\documentclass[twocolumn,showpacs,preprintnumbers,amsmath,amssymb]{revtex4}

\usepackage{graphicx}
\usepackage{epsfig}
\usepackage{dcolumn}
\usepackage{bm}

\begin{document}

\title {On the inequivalence of statistical ensembles}

\author{F. Gulminelli$^{(1)},$ Ph. Chomaz$^{(2)}$}

\affiliation{
(1) LPC Caen(IN2P3-CNRS/ISMRA et Universit\'{e}), F-14050 Caen c\'{e}dex,
France \\
(2) GANIL (DSM-CEA/IN2P3-CNRS), B.P.5027, F-14021 Caen c\'{e}dex, France}

\begin{abstract}
We investigate the relation between various statistical
ensembles of finite systems. If ensembles differ at the level of
fluctuations of the order parameter, we show that the equations of states
can present major differences. A sufficient
condition for this inequivalence to survive at the thermodynamical limit is
worked out. If energy consists in a kinetic and a
potential part, the microcanonical ensemble does not converge towards the
canonical ensemble when the partial heat capacities per particle fulfill the
relation $c_{k}^{-1}+c_{p}^{-1}<0$.
\end{abstract}


\maketitle

In most textbooks the equivalence between the different statistical
ensembles is 
demonstrated at the
thermodynamical limit through the Van Hove theorem \cite{huang}. Indeed
ensembles differ at the level of fluctuations which are generally believed
to induce small corrections in finite systems and to become negligeable at
the limit of infinite systems.

In this paper we will show that this might not be always the case. For
finite systems, two ensembles which put different constraints on the
fluctuations of the order parameter lead to qualitatively
different equations of
states close to a first order phase transition. As an example the
microcanonical heat capacity may diverge to become negative while the
canonical one remains always positive and finite \cite{gauss,gross}. Such
inequivalences may survive at the thermodynamical limit for systems
involving long range forces
\cite{lynden-bell,ruffo}. Looking at the general properties of the order
parameter distribution a sufficient condition for this behavior to show up
can be explitely worked out.

Let us first concentrate on finite systems. For simplicity we will consider
the microcanonical and the canonical ensemble characterized by the energy $E$
and the temperature $\beta^{-1}$ respectively, but our
discussion is valid for any couple of conjugated extensive and intensive
variables.

The microcanonical ensemble is characterized by the level density $W(E)$
and the entropy $S=\log W$ .The caloric curve
is then $T^{-1}=\partial _{E}S.$ The canonical partition sum is the Laplace
transform of $W$: $Z_{\beta }=\int dEW\exp (-\beta E)$. In this article
we will assume that the partition sum converges; this is not always the case
\cite{ang_mom} and indeed the impossibility to
normalize the distribution $W\exp (-\beta E)$ is already a known case of
ensemble inequivalence.

In finite systems, the canonical ensemble differs from the microcanonical
one since it does not correspond to a unique energy but to a distribution $%
P_{\beta }(E)=\exp (S(E)-\beta E-\log Z_{\beta })$. 
If $P_{\beta }$ has a single maximum the average
energy $\left\langle E\right\rangle _{\beta }=-\partial _{\beta }\ln
Z_{\beta }$ can also be computed using a saddle point approximation
around the most probable energy $\overline{E}_{\beta }$

\begin{equation}
\left\langle E\right\rangle _{\beta }=\int dE E P_{\beta
}(E) =  \int dE E 
e^{-\frac{(E-\overline{E}_{\beta })^{2}}{2C}}
\ g_{\beta }(E-\overline{E})
\label{EQ:_E_average}
\end{equation}

with $g_{\beta }(x)=c_{0}+c_{3}x^{3}+c_{4}x^{4}+\ldots $ . If $P_{\beta }$
is symmetric, $\left\langle E\right\rangle _{\beta }=\overline{E}_{\beta }$.
The definition of saddle implies 
\begin{equation}
T^{-1}\equiv \partial _{E}S\left( \overline{E}_{\beta }\right) =\beta
\label{EQ:_E_extremum}
\end{equation}%
meaning that the microcanonical caloric curve $T(\overline{E})$ exactly
coincides with the canonical one $\beta ^{-1}(\left\langle E\right\rangle ).$
However in a finite system the distribution may be not symmetric so that the two
curves can be shifted : $\left\langle E\right\rangle _{\beta }=%
\overline{E}_{\beta }+\delta _{\beta },$ where $\delta _{\beta
}=\int dx\ x\exp (-x^{2}/2C)\tilde{g}_{\beta }(x)=3c_{3}%
\sqrt{2\pi C^{5}}+...$ with $\tilde{g}_{\beta }$ the series of the odd terms
of $g_{\beta }$. However, the shift $\delta $ is in most cases small so that
when $P_{\beta }$ has a unique maximum the ensembles are
almost equivalent even for a finite system.

A more interesting situation occurs in first order phase transitions where  
$P_{\beta }$ has a characteristic bimodal shape \cite%
{binder,labastie,topology} with two maxima 
$\overline{E}_{\beta}^{\left( 1\right) }$,
$\overline{E}_{\beta}^{\left( 2\right) }$ that can be associated with the 
two phases and a minimum $\overline{E}^{\left( 0\right) }$. 
These three solutions of Eq.(\ref{EQ:_E_extremum}) 
imply a backbending for the microcanonical caloric curve.
A single saddle point approximation is not valid in this case; however
it is always possible to write $P_{\beta }=$ $m_{\beta }^{\left( 1\right)
}P_{\beta }^{\left( 1\right) }+m_{\beta }^{\left( 2\right) }P_{\beta
}^{\left( 2\right) }$ with $P_{\beta }^{\left( i\right) }$ mono-modal
normalized probability distribution peaked at $\overline{E}_{\beta }^{\left(
i\right) }$. The canonical mean energy is then the
weighted average of the two energies

\begin{equation}
\left\langle E\right\rangle _{\beta }=\tilde{m}_{\beta }^{\left( 1\right) }%
\overline{E}_{\beta }^{\left( 1\right) }+\tilde{m}_{\beta }^{\left( 2\right)
}\overline{E}_{\beta }^{\left( 2\right) }  \label{EQ:_E_can_backbend}
\end{equation}%
with $\tilde{m}_{\beta }^{\left( i\right) }=m_{\beta }^{\left( i\right)
}\int dE P_{\beta }^{\left( i\right) }(E)E/\overline{E}%
_{\beta }^{\left( i\right) }\simeq m_{\beta }^{\left( i\right) }$ , the last
equality holding for symmetric distributions $P_{\beta }^{\left( i\right) }$%
. As before correcting terms depending on the skewness $c_{3}^{\left(
i\right) }$ can be easily derived.

Since only one mean energy is associated with a given
temperature $\beta ^{-1},$ the canonical caloric curve is monotonous,
meaning that in the first order phase transition region
the two ensembles are not equivalent 

If instead of looking at the average $\left\langle E\right\rangle _{\beta }$
we look at the most probable energy $\overline{E}_{\beta }$ , this
(unusual) canonical caloric curve is identical to the microcanonical one
(see eq. (\ref{EQ:_E_extremum})) up to the transition temperature $\beta
_{t}^{-1}$ for which the two components of $P_{\beta }\left( E\right) $ have
the same height.
At this point the
most probable energy jumps from the low to the high energy branch of the
microcanonical caloric curve. The canonical curve is still monotonic 
and presents a
plateau at $\beta _{t}^{-1}$ which is equivalent to the Maxwell construction
since

\begin{equation}
S(\overline{E}_{\beta }^{\left( 2\right) })-S(\overline{E}_{\beta }^{\left(
1\right) })=\int_{\overline{E}_{\beta }^{\left( 1\right) }}^{\overline{E}%
_{\beta }^{\left( 2\right) }}\frac{dE}{T}=\beta \left( \overline{E}_{\beta
}^{\left( 2\right) }-\overline{E}_{\beta }^{\left( 1\right) }\right) 
\end{equation}

The question arises whether this violation of ensemble equivalence 
survives towards the thermodynamical limit. This
limit can be expressed as the fact that the thermodynamical
potentials per particle converge when the number of particles $N$ goes to
infinity : $f_{N,\beta }=\beta ^{-1}\log Z_{\beta }/N \rightarrow 
\bar{f}_{\beta }$ and $s_{N}\left( e\right)
=S(E)/N \rightarrow
\bar{s}\left( e\right) $
where $e=E/N$. Let us also introduce the reduced probability $p_{N,\beta
}\left( e\right) =\left( P_{\beta }(N,E)\right) ^{1/N}$ which then converges
towards an asymptotic distribution $p_{N,\beta }\left( e\right) \rightarrow
\bar{p}_{\beta }\left(e\right) $ where 
$\bar{p}_{\beta }\left( e\right) =\exp
\left( \bar{s}(e)-\beta e+\bar{f}_{\beta }\right) $. Since $P_{\beta
}(N,E)\approx $ $\left( \bar{p}_{\beta }\left( e\right) \right) ^{N}$ one
can see that when $\bar{p}_{\beta }\left( e\right) $ is normal the relative
energy fluctuation in $P_{\beta }(N,E)$ is suppressed by a factor $1/\sqrt{N}
$. At the thermodynamical limit $P_{\beta }$ reduces to a $\delta $-function
and the ensemble equivalence is recovered.

The situation is more complicated in the case of a first order phase
transition, i.e. for a bimodal $p_{N,\beta }\left( e\right) $ . As before,
let us introduce $\beta _{N,t}^{-1}$ the temperature for which the two
maxima of $p_{N,\beta }\left( e\right) $ have the same height. For a first
order phase transition $\beta _{N,t}^{-1}$ converges to a fixed point $\bar{%
\beta}_{t}^{-1}$ as well as the two maximum energies $e_{N,\beta }^{\left(
i\right) } \rightarrow   
\bar{e}_{\beta}^{\left( i\right) }$. 
For all temperature lower (higher) than $\bar{\beta}%
_{t}^{-1}$only the low (high) energy peak will survive at the
thermodynamical limit since the difference of the two maximum probabilities
will be raised
to the power $N.$ Therefore, below $\bar{e}_{\beta }^{\left( 1\right) }$ and
above $\bar{e}_{\beta }^{\left( 2\right) }$ the canonical caloric curve
coincides with the microcanonical one in the thermodynamical limit. In the
canonical ensemble the temperature $\bar{\beta}_{t}^{-1}$ corresponds to a
discontinuity in the state energy irrespectively of the behavior of
the entropy between $\bar{e}_{\beta }^{\left( 1\right) }$ and $\bar{e}%
_{\beta }^{\left( 2\right) }$.

\begin{figure}

\includegraphics*[height=0.5\linewidth,trim=100 250 100 50]{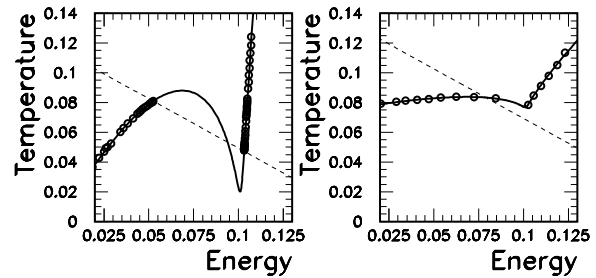}

\caption{Temperature as a function of
the potential energy $E_2$(full lines) and of the kinetic energy $E-E_2$
(dashed lines) for two model equation of states of classical systems showing
a first order phase transition. Symbols: temperatures extracted from the
most probable kinetic energy thermometer from eq.(5).}

\end{figure}

The microcanonical caloric curve in the phase transition region 
may either converge towards the Maxwell construction 
or keep a backbending behavior, since a negative heat capacity
system can be thermodynamically stable even in the thermodynamical limit if
it is isolated \cite{thirring}. This point has been recently made in
somewhat different words by Leyvraz and Ruffo\cite{leyvraz}. Examples of a
backbending behavior at the thermodynamical limit have been reported for a
model many-body interaction taken as a functional of the hypergeometric
radius in ref.\cite{lynden-bell} and for the long range Ising model
\cite{ruffo}. This can
be understood as a general effect of long range interactions for
which the topological anomaly leading to the convex intruder in the entropy
is not cured by increasing the number of particles\cite{ruffo,pettini}.
Conversely, for short range interactions \cite{gross} the backbending 
is a surface effect which should disappear at the
thermodynamical limit. This is the case for the microcanonical model of
fragmentation of atomic clusters\cite{gross-clus} and for the lattice gas
model with fluctuating volume\cite{europhys} . The interphase surface
entropy goes to zero as $N\rightarrow \infty $ in these models leading to a
linear increase of the entropy  in agreement with the canonical predictions.
From these examples, we can conclude that in the coexistence region the
microcanonical equation of states may remain different from the canonical
one even at the thermodynamical limit if the involved phenomena are not
reduced to short range effects. \ \ 

An especially interesting situation occurs for hamiltonians containing a
kinetic energy contribution:  if the kinetic heat capacity is large enough 
we will now show that the microcanonical curve presents at the
thermodynamical limit a temperature jump in complete disagreement with the
canonical ensemble.

Let us consider a finite system for which the hamiltonian can be separated
into two components $E=E_{1}+E_{2},$ that are statistically independent ($%
W(E_{1},E_{2})=W_{1}(E_{1})W_{2}(E_{2})$) and such that the associated 
degrees of freedom scale in the same way with the
number of particles; we will also consider the case where $S_{1}=\log W_{1}$
has no anomaly while $S_{2}=\log W_{2}$ presents a convex intruder which is
preserved at the thermodynamical limit. Typical examples of $E_{1}$ are
given by the kinetic energy for a classical system with velocity independent
interactions or other similar one body operators \cite{ruffo}.

\begin{figure}

\includegraphics*[height=0.5\linewidth,trim=100 250 100 50]{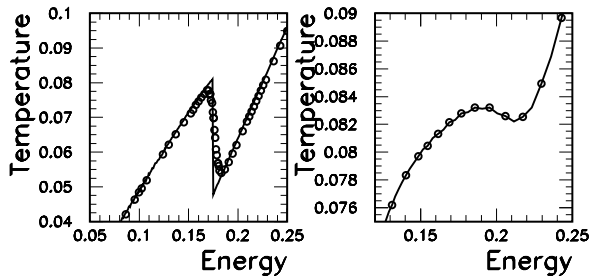}

\caption{Symbols: total caloric curves
obtained with the potential and kinetic equations of state of figure 1.
Lines: thermodynamical limit from a (double) saddle point approximation of
the partial energy microcanonical distributions.}

\end{figure}

The probability to get a partial energy $E_{1}$when the total energy is $E$
is given by

\begin{equation}
P_{E}\left( E_{1}\right) =\exp \left( S_{1}\left( E_{1}\right) +S_{2}\left(
E-E_{1}\right) -S\left( E\right) \right)  \label{eq:P1}
\end{equation}%
The extremum of $P_{E}\left( E_{1}\right) $ is obtained for the partitioning
of the total energy $E$ between the kinetic and potential components that
equalizes the two partial temperatures $\overline{T}=\partial _{E_{1}}S_{1}(%
\overline{E}_{1})=$ $\partial _{E_{2}}S_{2}(E-\overline{E}_{1})$. If $%
\overline{E}_{1}$is unique, $P_{E}\left( E_{1}\right) $ is mono-modal and we
can use a saddle point approximation around this solution to compute the
entropy $S\left( E\right) =\int_{-\infty }^{E}dE_{1}\exp \left( S_{1}\left(
E_{1}\right) +S_{2}\left( E-E_{1}\right) \right) .$ At the lowest order this
leads to the microcanonical temperature of the global system $\partial
_{E}S(E)=\overline{T}^{-1}$ meaning that the most probable partial energy $%
\overline{E}_{1}$acts as a microcanonical thermometer. If $%
\overline{E}_{1}$is always unique, the kinetic thermometer in the
backbending region will follow the whole decrease of temperature as the
total energy increases. Therefore, the total caloric curve will present the
same anomaly as the potential one.

If conversely the partial energy distribution is double humped \cite{berry}, 
then the
equality of the partial temperatures admits three solutions one of them $%
\overline{E}_{1}^{\left( 0\right) }$ being a minimum. At this point the
partial heat capacities $C_{1}^{-1}=-\overline{T}^{2}\,\partial
_{E_{1}}^{2}S_{1}(\overline{E}_{1}^{(0)})$ and $C_{2}^{-1}=-\overline{T}%
^{2}\,\partial _{E_{2}}^{2}S_{2}(E-\overline{E}_{1}^{(0)})$ fulfill the
relation 
\begin{equation}
C_{1}^{-1}+C_{2}^{-1}<0  \label{eqcentral}
\end{equation}%
This happens when the potential heat capacity is negative and the kinetic
energy has a sufficient number of degrees of freedom ($C_{1}>-C_{2}$) to act
as an approximate heat bath: the partial energy distribution $P_{E}\left(
E_{1}\right) $ in the microcanonical ensemble is then bimodal as the total
energy distribution $P_{\beta }\left( E\right) $ in the canonical ensemble.
In this case the microcanonical temperature is given by a weighted average
of the two estimations from the two maxima of the kinetic energy
distribution 
\begin{equation}
T=\partial _{E}S(E)=\frac{\overline{P}^{(1)}\sigma ^{(1)}/\overline{T}^{\left(
1\right) }+\overline{P}^{(2)}\sigma ^{(2)}/\overline{T}^{\left( 2\right) }}{%
\overline{P}^{(1)}\sigma ^{(1)}+\overline{P}^{(2)}\sigma ^{(2)}}
\label{zero_princ}
\end{equation}%
where $\overline{T}^{\left( i\right) }=T_{1}(\overline{E}_{1}^{\left(
i\right) })$ are the kinetic temperatures calculated at the two maxima, $%
\overline{P}^{\left( i\right) }=P_{E}(\overline{E}_{1}^{\left( i\right) })$
are the probabilities of the two peaks and $\sigma ^{\left( i\right) }$
their widths. At the thermodynamical limit eq.(\ref{eqcentral}) reads $%
c_{1}^{-1}+c_{2}^{-1}<0$, with $c=\lim_{N \to \infty }C/N$
. If this condition is fulfilled the probability distribution 
$P_{\beta}(E)$ presents two
maxima for all finite sizes and only the highest peak survives at $N=\infty $%
. Let $E_{t}$ be the energy at which $P_{E_{t}}(\overline{E}^{\left(
1\right) })=P_{E_{t}}(\overline{E}^{\left( 2\right) })$. Because of the zero
principle eq.(\ref{zero_princ}) at the thermodynamical limit the caloric
curve will follow the high (low) energy maximum of $P_{E}\left( E_{1}\right) 
$ for all energies below (above) $E_{t}$; there will be a temperature jump
at the transition energy $E_{t}.$

Let us illustrate the above results with two examples for a classical gas of
interacting particles. For the kinetic energy contribution we have $%
S_{1}(E)=c_{1}\ln (E/N)^{N}$ with a constant kinetic heat capacity per
particle $c_{1}=3/2$. For the potential part we will take two polynomial
parametrizations of the interaction caloric curve presenting a back bending
which are displayed in figure 1. If the decrease of the partial temperature $%
T_{2}(E_{2})$ is steeper than $-2/3$ (left part) \cite{lynden-bell}
eq.(\ref{eqcentral}) is verified and the kinetic caloric curve $T_1(E-E_1)$
(dashed line) crosses the potential
one $T_2(E_2)$ (full line) in three different points for all values of the total energy
lying inside the coexistence region. The resulting caloric curve for the
whole system is shown in figure 2 (symbols) together with the
thermodynamical limit (lines) evaluated from the double saddle point
approximation (\ref{zero_princ}). In this case one observes a temperature
jump at the transition energy while if the temperature decrease is smoother
(right part of figures 1 and 2) the shape of the interaction caloric curve
is preserved at the thermodynamical limit.

\begin{figure}

\includegraphics*[height=0.5\linewidth,trim=300 100 300 100]{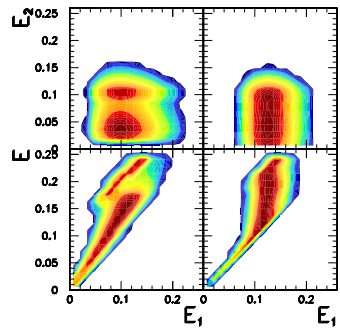}

\caption{Canonical event distributions
in the potential versus kinetic energy plane (upper part) and total versus
kinetic energy plane (lower part)  at the transition temperature for the two
model equations of state of figure 1. }

\end{figure}

The occurrence of a temperature jump in the thermodynamical limit is easily
spotted by looking at the bidimensional canonical event distribution $%
P_{\beta }(E_{1},E_{2})$ in the partial energies plane. This density of
states is just the product of the independent kinetic and potential
canonical probabilities as shown in the upper part of figure 3 for the two
model equation of states of figure 1 at the transition temperature $\beta
=\beta _{t}$. In order to discuss the microcanonical ensemble one has to
introduce the total energy $E=E_{1}+E_{2}.$ Thus we can look at the
canonical distribution as a function of $E$ and $E_{1}$ 
\begin{equation}
P_{\beta }(E,E_{1})\propto\exp S_{1}(E_{1})\exp S_{2}(E-E_{1})\exp (-\beta E)
\end{equation}%
which is shown in the lower part of figure 3. The deformation of the event
distribution induced by the microcanonical constraint does not cause a
topological difference between our two model cases; this explains why both
converge to the Maxwell construction for $N\rightarrow \infty $ in the
canonical ensemble. If we now study the microcanonical ensemble we have to
look at constant energy cuts of $P_{\beta }(E,E_{1})$ leading to the
microcanonical distribution $P_{E}(E_{1})$ within a normalization constant.
If the anomaly in the potential equation of state is sufficiently important,
the distortion of events is such that one can still see the two phases
coexist even after a sorting in energy. Figure 4 shows two cuts of the lower
part of figure 3 at an energy close to the transition energy. In the system
which does not present a temperature jump the partial energy distribution is
normal, while the two peaks of the system violating ensemble equivalence can
be interpreted as the precursors of phases \cite{topology} . This can also
be seen from the most probable kinetic temperature (symbols in the left part
of figure 1) which makes a sudden jump at the transition energy.

\begin{figure}

\includegraphics*[height=0.5\linewidth,trim=70 70 70 50]{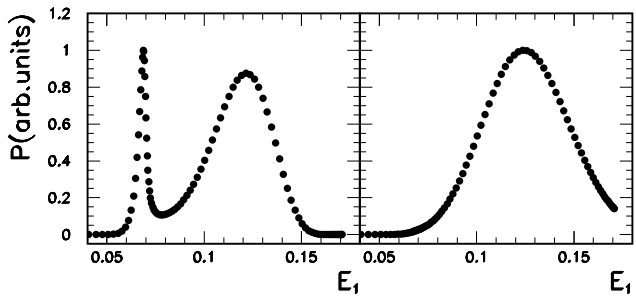}

\caption{ Microcanonical kinetic energy
distributions inside the coexistence region for the two model equations of
state of figure 1.}

\end{figure}

\smallskip

In conclusion, in this paper we have analyzed the distribution of observable
quantities related to the order parameter in finite systems undergoing a
first order phase transition. This phenomenon is uniquely signed by a
bimodality of the probability distribution in the intensive ensemble where
the Lagrange parameter associated to the considered observable is fixed. In
such a physical situation the different statistical ensembles are not in
general equivalent even at the thermodynamical limit. In finite systems, the
order parameters
(e.g. energy in the
considered examples) have an average value which varies smoothly while the
most probable value makes a jump as a function of the associated Lagrange
multiplier (e.g. temperature) in the intensive (e.g. canonical) ensemble. In
the corresponding extensive (e.g. microcanonical) ensemble the corresponding
equation of states (e.g. the caloric curve) presents a back bending. In
infinite systems, this inequivalence between statistical ensembles 
may remain. We have shown
that a generic behavior of the extensive ensemble can be a discontinuity in
the associated intensive variable at a given value of the fixed extensive
variable. The condition for this ensemble inequivalence can be explicitly
worked out in a wide range of physical systems. In particular,
microcanonical caloric curves present a temperature jump at the
thermodynamical limit if the negative heat capacity is sufficiently small in
absolute value for the kinetic energy to play the role of a heat bath.


\end{document}